\input harvmac.tex
%
%
\def\eps{\epsilon}
\def\epb{\bar\epsilon}
\def\psb{\bar\psi}
\def\lb{\bar\lambda}
\def\fb{\bar\Phi}
\def\si{\sigma^i}
\def\sj{\sigma^j}
\def\scrn{{\cal{N}}}
\def\donenfour{{d=1,\, \scrn=4}}
\def\doneneight{{d=1,\, \scrn=8}}
\def\dfournone{d=4,\, \scrn=1}
\def\dthreentwo{d=3,\, \scrn=2}

\def\kii{K_{\Sigma^i\Sigma^i}}
\def\klli{K_{\Sigma^l\Sigma^l\Sigma^i}}
\def\kllk{K_{\Sigma^l\Sigma^l\Sigma^k}}
\def\kiif{K_{\Sigma^i\Sigma^i\Phi}}
\def\kiib{K_{\Sigma^i\Sigma^i\fb}}
\def\kllii{K_{\Sigma^l\Sigma^l\Sigma^i\Sigma^i}}

\def\kiijf{K_{\Sigma^i\Sigma^i\Sigma^j\Phi}}
\def\kiijb{K_{\Sigma^i\Sigma^i\Sigma^j\fb}}
\def\kiiff{K_{\Sigma^i\Sigma^i\Phi\Phi}}
\def\kiibb{K_{\Sigma^i\Sigma^i\fb\fb}}

\def\kff{K_{\Phi\fb}}
\def\kffi{K_{\Phi\fb\Sigma^i}}
\def\kffk{K_{\Phi\fb\Sigma^ik}}
\def\kfff{K_{\Phi\fb\Phi}}
\def\kffb{K_{\Phi\fb\fb}}
\def\kffff{K_{\Phi\fb\Phi\Phi}}
\def\kffff{K_{\Phi\fb\fb\fb}}
\def\kffif{K_{\Phi\fb\Sigma^i\Phi}}
\def\kffib{K_{\Phi\fb\Sigma^i\fb}}
\def\kffij{K_{\Phi\fb\Sigma^i\Sigma^j}}

\def\bart{{\bar\theta}}
\def\covda{{\partial\over{\partial{\theta}^{\alpha}}}-i\bart_{\alpha}
{\partial}_0}
\def\covdbara{{\partial\over{\partial{\bar{\theta}}^{\alpha}}}-i{\theta}_{
\alpha}{\partial}_0}
\def\qa{{\partial\over{\partial{\theta}^{\alpha}}}+i{\bar{\theta}}_{\alpha}
{\partial}_0}
\def\qbara{{\partial\over{\partial{\bar{\theta}}^{\alpha}}}+i{\theta}_{
\alpha}{\partial}_0}
\def\sig{\Sigma_{\alpha\beta}}
\def\fourtheta{\theta\theta{\mathop{\overline{\theta\theta}}}}
\def\twotheta{\mathop{\overline{\theta\theta}}}
\def\parz{\partial_0}

\def\np#1#2#3{{\it{Nucl. Phys.}} {\bf{B{#1}}} (#2) #3}
\def\pl#1#2#3{{\it{Phys. Lett.}} {\bf{B{#1}}} (#2) #3}
\def\prd#1#2#3{{\it{Phys. Rev.}} {\bf{D{#1}}} #2 (#3)}

\nref\DKPS{M. R. Douglas, D. Kabat, P. Pouliot and S. Shenker, {\it
D-Branes and Short Distances in String Theory},
\np{485}{1997}{85}, hep-th/9608024.}
\nref\BFSS{T. Banks, W. Fischler, S. H. Shenker and L. Susskind, {\it
M-theory as A Matrix Model: A Conjecture.}
\prd{55}{1997}{5112}, hep-th/9610043.}
\nref\MD{M.R. Douglas, {\it  Superstring Dualities, Dirichlet Branes
and the Small Scale Structure of Space}, Proceedings of the LXIV Les
Houches session on `Quantum Symmetries', August 1995, hep-th/9610041.}
\nref\GHR{S.J. Gates, Jr., C.M. Hull and M. Ro\v{c}ek, {\it Twisted Multiplets
and New Supersymmetric Non-linear $\sigma$-models},
\np{248}{1984}{157}.}
\nref\WITC{E. Witten, {\it Phases of $N=2$ Theories In Two
Dimensions}, \np{403}{1993}{159}, hep-th/9301042.}
\nref\POLD{S.~Chaudhuri, C.~Johnson and J.~Polchinski,
{\it Notes on D-Branes}, hep-th/9602052.}
\nref\TASI{J. Polchinski, {\it TASI Lectures on D-Branes}, hep-th/9611050.} 
\nref\SEIB{N. Seiberg, {\it IR Dynamics on Branes and Space-Time
Geometry}, \pl{384}{1996}{81}, hep-th/9606017.}
\nref\SW{N. Seiberg and E. Witten, {\it Gauge Dynamics and
Compactification to Three Dimensions}, hep-th/9607163.}
\nref\IS{K. Intriligator and N. Seiberg, {\it Mirror Symmetry in Three
Dimensional Gauge Theories}, \pl{387}{1996}{513}, 
hep-th/9607207.}
\nref\HAR{J. Harvey, {\it Jerusalem Winter School on Strings and
Duality Lectures.}}


\Title{\vbox{\baselineskip12pt\hbox{hep-th/9706059}
\hbox{RU-97-50}}}
{\vbox{
\centerline{A Non-Renormalization Theorem For}
\vskip 10pt
\centerline{The $\doneneight$ Vector Multiplet}
}}
\centerline{Duiliu-Emanuel Diaconescu and Rami Entin}
\medskip
\centerline{\it Department of Physics and Astronomy}
\centerline{\it Rutgers University }
\centerline{\it Piscataway, NJ 08855--0849}
\medskip
\centerline{\tt duiliu, rami@physics.rutgers.edu}
\medskip
\bigskip
\noindent

Sigma models describing low energy effective actions on D0-brane
probes with $\scrn=8$ supercharges are studied in detail using 
a manifestly $\donenfour$ super-space formalism. Two $0+1$ 
dimensional $\scrn=4$ multiplets together with their general actions
are constructed. We derive the condition for these actions to
be $\scrn=8$ supersymmetric and apply these techniques to various
D-brane configurations. We find that if in addition to $\scrn=8$
supersymmetry the action must also have $Spin(5)$ invariance, the form
of the sigma model metric is uniquely determined by the one-loop
result and is not renormalized perturbatively or non-perturbatively.

\Date{June 1997}

\newsec{Introduction and summary}

Recent developments \refs{\DKPS, \BFSS, \MD} have emphasized the
crucial role played by D0-branes in probing space-time structure at
sub-stringy scales as well as in a non-perturbative definition of
eleven dimensional M-theory. The basic feature that enables
D-particles to test short distances in string theory is that their low
energy dynamics is a quantum mechanics of the lightest open string
degrees of freedom. The geometrical background in the sub-stringy
domain is reproduced by quantum open string effects while the
classical background at distances larger than the string scale is
described by supergravity results which are essentially mediated by
massless closed strings. As discussed in \DKPS , in the cases with
enough supersymmetry the two regimes are continuously connected by
factorization of the open string annulus diagram. The general behavior
in such cases is that the long distance supergravity results coincide
with the the one-loop quantum corrections to the probe moduli space. 
By analogy with higher dimensional field theories it is plausible that
higher order perturbative corrections as well as non-perturbative ones  
vanish, leading to non-renormalization theorems.
A similar non-renormalization result for a higher derivative
interaction proves to be essential \BFSS\ in the Matrix theory formulation
of M-theory.

The purpose of the present work is to study
the $\scrn=8$ quantum mechanics of a D0-brane probe moving in
different D4-brane and/or orientifold plane backgrounds. The low
energy degrees of freedom in the probe theory are five bosons and
eight fermions. A single D0-D4 configuration has $\scrn=8$
supersymmetry and a $Spin(5)$ rotational symmetry in the transverse
directions under which the bosons transform as a vector and the fermions as
a spinor. We construct two $\scrn=4$ multiplets that together have
these degrees of freedom, but are not manifestly $Spin(5)$ symmetric.
We call the pair of these multiplets the $\doneneight$ vector
multiplet. Our main result is that the condition for $Spin(5)$
invariance of the vector multiplet action is compatible with the
condition for it to have $\scrn=8$ supersymmetry, and that when taken
together these invariances uniquely determine the form of the target
space metric. The form of the metric we find agrees with the one-loop
result of \DKPS\ and we conclude that it can not receive perturbative
or non perturbative corrections.

The plan of this paper is as follows. In section 2 we develop a
manifestly $\scrn=4$ super-space formalism in $(0+1)$ dimensions and
describe $\scrn=4$ chiral and linear multiplets that together contain
the right number of bosonic and fermionic degrees of freedom. We then
find in section 3 the condition for this action to admit four
additional supersymmetries and argue that this condition is
essentially unique. Requiring also $Spin(5)$ invariance leads to the
non-renormalization theorem. This result is applied in section 4 to
various D-brane configurations. Finally, we discuss the range of
validity of this theorem in connection with three dimensional
analogues and string duality.

\newsec{$\scrn=4$ multiplets in one dimension}

The $\donenfour$ superspace is
parameterized by one commuting coordinate $t$, and
four non-commuting ones arranged as an $SU(2)$ spinor, $\theta_\alpha$ 
and its complex conjugate $\bart^\alpha$. The covariant derivatives 
and supercharges are given by (our conventions are summarized in
appendix A):
$$\eqalign{D_{\alpha}&=\covda\hskip 20pt\bar{D}_{\alpha}=\covdbara\cr
Q_{\alpha}&=\qa\hskip 20pt\bar{Q}_{\alpha}=\qbara ,}$$
and satisfy the algebra
\eqn\algone{\{D_{\alpha},{\bar{D}}_{\beta}\}=2i{\eps}_{\alpha\beta}\parz
\hskip 20pt
\{Q_{\alpha},{\bar{Q}}_{\beta}\}=-2i{\eps}_{\alpha\beta}\parz ,}
with all the other anticommutators vanishing. The manifest supersymmetry
transformations are generated by
$\eps^\alpha Q_\alpha+{\epb}^\alpha{\bar{Q}}_\alpha$
acting on the various multiplets.
\vskip 5pt
{\it{Chiral Multiplet}}
\vskip 5pt
As in the $\dfournone$ case, the chiral and antichiral multiplets are
defined by the constraints ${\bar{D}}\Phi=D{\fb}=0$, 
which are solved by functions of $y=t-i\theta^\alpha\bart_\alpha$ and 
${\bar{y}}=t+i\theta^\alpha\bart_\alpha$.
In component form  they are given by
$$\eqalign{\Phi(y)&=\Phi(y)+2\theta^{\alpha}\psi_{\alpha}(y)
+\theta\theta F(y)\cr
&=\Phi-i\theta^{\alpha}{\bar{\theta}}_{\alpha}\dot{\Phi}
+{1\over{4}}\fourtheta{\ddot{\phi}}
+2\theta^{\alpha}\psi_{\alpha}
-i\theta\theta{\bar{\theta}}_{\alpha}{\dot{\psi}}^{\alpha}
+\theta\theta F}$$
and
$$\eqalign{\qquad\,\,{\fb}({\bar{y}})&={\fb}({\bar{y}})
-2{\bar{\theta}}_{\alpha}{\psb}^{\alpha}({\bar{y}})
-\twotheta F^*({\bar{y}})\cr
&={\fb}+i\theta^{\alpha}\bart_{\alpha}{\dot{\fb}}
+{1\over{4}}\fourtheta{\ddot{\fb}}
-2{\bar{\theta}}_{\alpha}{\psb}^{\alpha}
+i\twotheta\theta_{\alpha}{\dot{\psb}}^{\alpha}
-\twotheta F^*}$$
which are the $\dfournone$ chiral and antichiral multiplets
reduced to one dimension. The physical on-shell degrees of freedom arising
from these multiplets are two bosons and four fermions.
\vskip 5pt
{\it{Linear Multiplet}}
\vskip 5pt
The $\dfournone$ vector multiplet dimensionally reduced to $D=3$ becomes
equivalent \GHR\ to the real linear multiplet $G$ defined by the
constraints
$$D^2 G={\bar{D}}^2 G=0,$$
where $D,\,\bar D$ denote the spinor derivatives of the $\dthreentwo$
superspace. They are solved by
$$
G=D{\bar D}V
$$ 
with $V$ an arbitrary real superfield. The physical degrees of freedom
consist of a real scalar boson, a three dimensional vector field 
and their fermionic superpartners. The real scalar can be thought of
as the fourth component of the four dimensional vector field.
Further reduction to two dimensions yields \refs{\GHR ,\WITC} the
twisted chiral multiplet $\Sigma_{+-}$ defined \GHR\ by the constraints
$$
\bar D_{+}\Sigma_{+-}=D_{-}\Sigma_{+-}=0.
$$
The solution of these constraints can be expressed similarly in terms
of a real superfield
$$
\Sigma_{+-}={1\over {\sqrt 2}} \bar D_+ D_-V,
$$
describing the dynamics of the two real scalars obtained by
dimensional reduction of the four dimensional vector field plus their
superpartners.

In one dimension the closest analogue of the above conditions would
be
\eqn\constone{D^2\Sigma=\bar D^2\Sigma=0.}
Proceeding naively we take
$$
\Sigma=\bar{D}_\alpha\Theta^\alpha
$$
where $\Theta^\alpha$ is a superfield, as the general solution of the
second constraint in \constone . By making use of
$[\bar{D}_\alpha,D^2]=4iD_\alpha\parz$ the first condition in \constone\
is satisfied if 
$$
\bar{D}_\alpha D^2\Theta^\alpha+4iD_\alpha\parz\Theta^\alpha=0
$$
which is unacceptable since the time dependence of $\Theta^\alpha$ is
restricted. A natural modification would be to consider a triplet of
superfields which we denote by $\sig$. More precisely, the linear
multiplet $\sig$ is defined by 
\eqn\consttwo{D^{\gamma}D^{\alpha}\Sigma_{\alpha\beta}={\bar{D}}^{\gamma}{\bar{D}}^{\alpha}\Sigma_{\alpha\beta}=0,}
and the reality condition
\eqn\realone{{\bar{\Sigma}}_{\alpha\beta}\equiv\Sigma^{\alpha\beta}=\eps^{\alpha\gamma}\Sigma_{\gamma\delta}\eps^{\delta\beta}.}
The unique solution of these constraints with no restriction on the time
dependence is given by
\eqn\linone{\Sigma_{\alpha\beta}={\bar{D}}_{(\alpha}D_{\beta)}V,}
where $V$ is the real superfield of $\dfournone$ reduced to one
dimension and $()$ denote symmetrization. 
The second constraint in \consttwo\ is identically satisfied,
while the first one follows from the algebra \algone :
\eqn\norest{D^2\Sigma_{\alpha\beta}={1\over{2}}(D^2{\bar{D}}_{\alpha}D_{\beta}+D^2{\bar{D}}_{\beta}D_{\alpha})V
=i(\eps_{\beta\alpha}+\eps_{\alpha\beta})\parz V.}
The component form of $\sig$ is given by
\eqn\lintwo{\eqalign{\Sigma_{\alpha\beta}=&-{\sigma}^i_{\alpha\beta}x_i
+i(\theta_{\alpha}{\lb}_{\beta}+\theta_{\beta}{\lb}_{\alpha})
+i({\bar{\theta}}_{\beta}\lambda_{\alpha}+{\bar{\theta}}_{\alpha}\lambda_{\beta})
-({\bar{\theta}}_{\beta}\theta_{\alpha}+{\bar{\theta}}_{\alpha}\theta_{\beta})D\cr
&+{i\over{2}}(\theta_{\beta}{\bar{\theta}}^{\gamma}{\sigma}^i_{\gamma\alpha}+
\theta_{\alpha}{\bar{\theta}}^{\gamma}{\sigma}^i_{\gamma\beta}+
{\bar{\theta}}_{\beta}\theta^{\gamma}{\sigma}^i_{\gamma\alpha}+
{\bar{\theta}}_{\alpha}\theta^{\gamma}{\sigma}^i_{\gamma\beta}){\dot{x}}_i\cr
&+{1\over{2}}\twotheta(\theta_{\beta}{\dot{\lambda}}_{\alpha}+
\theta_{\alpha}{\dot{\lambda}}_{\beta})
-{1\over{2}}\theta\theta({\bar{\theta}}_{\beta}{\dot{\lb}}_{\alpha}+
{\bar{\theta}}_{\alpha}{\dot{\lb}}_{\beta})
+{1\over{4}}\fourtheta{\sigma}^i_{\alpha\beta}{\ddot{x}}_i
,}}
or alternatively by
\eqn\linthree{\eqalign{\Sigma^i\equiv&{1\over{2}}\sigma^{i\,\alpha\beta}\sig\cr
=&-x^i+i\theta_{\gamma}\sigma^{i\,\gamma\delta}{\lb}_{\delta}+
i\bart_{\gamma}\sigma^{i\,\gamma\delta}\lambda_{\delta}-\bart_{\gamma}
\sigma^{i\,\gamma\delta}\theta_{\delta}D\cr
&+\eps^{ijk}\bart_{\gamma}\sigma^{j\,\gamma\delta}\theta_{\delta}{\dot{x}}^k
-{1\over{2}}\twotheta\theta_{\gamma}\sigma^{i\,\gamma\delta}{\dot{\lambda}}_
{\delta}+{1\over{2}}\theta\theta\bart_{\gamma}\sigma^{i\,\gamma\delta}
{\dot{\lb}}_{\delta}+{1\over{4}}\fourtheta{\ddot{x}}^i}}
which is more convenient for our purpose. A supersymmetric lagrangian
will be a general function of the $\Sigma^i$ 
superfields\footnote{$^1$}{In the original version of the paper we 
erroneously stated that a general superspace action density must be an $SO(3)$ invariant function. We thank E. Witten for pointing out this 
mistake.}

Finally we note that by analogy with the $\dfournone$ case one can
define chiral and antichiral field-strength multiplets by
\eqn\fieldstrength{W_{\alpha}\equiv {\bar{D}}^{\beta}\sig\hskip 20pt
{\bar{W}}_{\alpha}\equiv D^{\beta}\sig.}
Then,
$$-{1\over{6}}\left(W^{\alpha}W_{\alpha}|_{\theta\theta}
+{\bar{W}}^{\alpha}{\bar{W}}_{\alpha}|_{\twotheta}\right)$$
yields the same kinetic terms as $\Sigma^i\Sigma^i|_{\fourtheta}$.

\newsec{$\scrn=8$ supersymmetry and non-renormalization}

We will ultimately be concerned with applications to a D0-brane in
D4-branes and orientifold backgrounds. As will be explained in section
4, the low energy degrees of freedom on the D0-brane world-line are
precisely described by the pair $(\Phi,\Sigma)$ which is the
$\doneneight$ multiplet. Such systems have only eight supersymmetries
so quadratic terms in the velocities are generally not protected from
renormalization. In the regime where the velocity of the D0-brane is
small we may restrict our attention to an action which is quadratic in
velocities and neglect higher order terms. A general such action with
four manifest supersymmetries is given by
\eqn\lag{\int d^2\theta d^2\bart\,K(\Phi,{\fb},\Sigma^i)}
where $K$ is an arbitrary real prepotential. It is possible to add a
superpotential integrated over half of superspace, but it will not contribute
to the metric. This remark actually applies to a wider class of actions. We
may think of \lag\ as the first term in an expansion of the form
$$
L=\int d^2\theta d
^2\bart
\left(K_2+K_{4ij}\parz\Sigma^i\parz\Sigma^j+
\tilde{K}_4\parz\Phi\parz\fb
+\ldots\right),
$$
where each successive $K_i$ produces an $i$-th power in velocity term in the
lagrangian (we do not have to consider an expansion in covariant
derivatives of the multiplets since these lead to cubic terms in the
velocities). Again, $K_4,\,\tilde{K}_4,\ldots$ can not give metric
terms, so the non-renormalization result we will prove below applies
also to the metric terms in these actions as well.

The metric can be read from the kinetic terms arising from the
superspace integration,
\eqn\kinetic{{1\over{4}}\kii({\dot{x}}^j{\dot{x}}^j
+i({\lb}{\dot{\lambda}}+\lambda{\dot{\lb}}))
-K_{\Phi{\fb}}({\dot{\Phi}}{\dot{\fb}
+i({\psb}{\dot{\psi}}+\psi{\dot{\psb}}))},}
and consists of only two undetermined functions,
${1\over{4}}\kii$ (summation over $i$ is implicit here) and $-\kff$.
Note the absence of mixed derivative terms -- this will prove crucial
for the applications to the $0-4$ system.

\subsec{Non-manifest supersymmetries}

If the action \lag\ admits more supersymmetries their form is severely
constrained by the following considerations. First, they must be
realized as spinorial derivatives acting on superfields so that the
supersymmetry algebra is satisfied. The  four manifest 
supersymmetries of each multiplet are already generated by the supercharges
acting on it. Therefore, additional supersymmetries, if they exist, 
must be generated by spinorial derivatives acting on the other
multiplets. This means that $\sig$ will enter the non-manifest
transformations of $\Phi$, $\fb$ and that $\Phi$, $\fb$ enter
symmetrically the non-manifest variation of $\sig$. The form of these
variations is further constrained, and in fact determined up to a
constant, by requiring the variations to respect the defining constraints
of the multiplets. Thus we conclude that if there are four additional
supersymmetries their form is
\eqn\nonmanone{\eqalign{\delta{\Phi}&\propto i{\epb}^{\beta}{\bar{D}}^{\alpha}\sig\cr
\delta{\fb}&\propto i\eps^{\beta}D^{\alpha}\sig\cr
\delta\sig&\propto i(\eps_{(\alpha}D_{\beta)}\Phi-{\epb}_{(\alpha}{\bar{D}}_{\beta)}\fb)}}
(The chiral and antichiral constraints of $\delta\Phi$ and $\delta\fb$
follow directly form \consttwo . The variation of $\sig$ can be seen to
satisy the conditions \realone\ and \consttwo\ by using the algbra
\algone . It is also easy to verify that the commutator of two
non-manifest variations closes on translations).

A straight-forward (and a little laborious) calculation in components shows
that the action \lag\ admits the four non-manifest supersymmetries
\eqn\nonmantwo{\eqalign{\delta{\Phi}&={-2i\over{3}}{\epb}^{\beta}{\bar{D}}^{\alpha}\sig\cr
\delta{\fb}&={-2i\over{3}}\eps^{\beta}D^{\alpha}\sig\cr
\delta\sig&=i(\eps_{(\alpha}D_{\beta)}\Phi-{\epb}_{(\alpha}{\bar{D}}_{\beta)}\fb),}}
provided that the following condition holds:
\eqn\punchline{\kii+4K_{\Phi\bar\Phi}=0.}
This is actually also a {\it necessary} condition. The explicit form of the
action shows that it cannot be invariant under the supersymmetry
transformations of the form \nonmanone\ unless \punchline\ holds, that
is unless the action depends only one arbitrary function. As
argued above, the form of the non-manifest variations is unique so we
conclude that any ${\scrn=4}$ supersymmetric action is
automatically $\scrn=8$ supersymmetric if and only if \punchline\ holds.
The metric, the action and all the supersymmetry transformations are
now determined by
\eqn\metone{f=-{1\over{4}}\kii=\kff}
and are given in appendix B ($f$ enters the variation laws once the
auxiliary fields are solved for). Differentiating $f$ twice with respect to
$\Sigma^i$ and with respect to $\Phi$ and $\fb$, and using \punchline
, shows that the metric satisfies
\eqn\mettwo{f_{\Sigma^i\Sigma^i}+4f_{\Phi\bar\Phi}=0}
as well.

\subsec{$Spin(5)$ invariance and non-renormalization}

The five scalars in the vector multiplet can be thought of as local
coordinates, $y_1,\ldots,y_5$, on a five dimensional target space
manifold by making the change of variables
$$
\eqalign{y_i&=x^i\hskip 20pt i=1,2,3\cr
y_4&={1\over{2}}(\Phi+\fb)\cr
y_5&={1\over{2i}}(\Phi-\fb).}
$$
In these coordinates the condition \mettwo\ satisfied by the metric
$f$ is precisely the $Spin(5)$ invariant Laplace equation. Any function of
$r^2$, $r$ being the five dimensional radius, must also satisfy this
equation, and therefore the condition for a $Spin(5)$ invariant
metric is compatible with the condition for $\scrn=8$
supersymmetry. This conclusion depends crucially on the relative sign
and factor in \kinetic\ and would not have been valid
otherwise. Furthermore, $f$ is now determined up to two constants. The
condition \mettwo\ on a $Spin(5)$ invariant function reduces to
\eqn\diffeq{r^2f^{\prime\prime}+{5\over{2}}f^{\prime}=0,}
and is solved by
\eqn\sol{f=C^{\prime}+{C\over r^3}}
where $C,\,C^\prime$ are arbitrary constants. We conclude that the
metric of a general action compatible with the above symmetries is not
renormalized either perturbatively or non-perturbatively.

It is also possible to restore manifest $Spin(5)$ invariance in the
full lagrangian. After solving algebraically for the auxiliary fields
the superspace lagrangian is given by \footnote{$^*$}{To avoid
clutter, the same tangent space indices are used to denote the flat
space carried by the $\gamma$ matrices.}
\eqn\fivelag{-f\left(\dot x^i\dot x^i +i\left(\bar\eta{\dot\eta}+
\eta\dot{\bar\eta}\right)\right)+{\dot x}^if_{,j}
\left(\eta\gamma^{ij}\bar\eta\right)+{1\over2}
\left(f_{,ij}-{1\over 2}f_{,i}f_{,j}\right)\left(\eta\gamma^i\bar\eta\,
\eta\gamma^j\bar\eta+\eta\gamma^i\eta\,\bar\eta\gamma^j\bar\eta\right),}
and has a natural geometric interpretation. Specifically the bilinear
fermion term, taking into account the f\"unfbein factors, is the
pull back of the minimal spin connection made of the metric $f$. With
a little algebra, the quadratic fermion term can also be seen to be
$$
R_{\alpha\beta\gamma\delta}\,\eta^{\alpha}\bar\eta^{\beta}\eta^{\gamma}\bar{\eta}^{\delta},
$$
with the curvature computed from the minimal connection. The manifest
and non-manifest SUSY transformations also match up in a nice way.
The manifest ones can be recovered if in the five dimensional SUSY
transformations (B.6), the parameter $\epsilon^5_\alpha$ is taken as 
$$
\eps^5_\alpha=\left(\matrix{\epsilon_\alpha\cr0}\right),
$$
and the non-manifest ones if we take 
$$
\eps^5_\alpha=\left(\matrix{0\cr\epsilon_\alpha}\right).
$$
Since the target space is odd dimensional these restrictions can not be made
in an invariant way, but together they combine into an $\scrn=8$ SUSY
parameter. This is again due to the consistency of the $Spin(5)$
invariance and $\scrn=8$ conditions.

\newsec{D0-D4 system}

The formalism developed in the previous sections can be applied to the
study of low energy effective actions of D0-brane probes in different
Type IIA backgrounds. Extending the analysis of \DKPS\ we consider 
D0-brane probes in Type I' theory realized as an orientifold of the 
Type IIA theory compactified on a five torus $T^5$ \refs{\POLD,
\TASI}. More precisely, one starts with Type I theory on $T^5$ and
performs T-duality on all the five circles of the torus . The
resulting theory is Type IIA on $T^5/Z_2\Omega$ with sixteen pairs of 
D4-branes in the background to cancel the charge of the 32 orientifold fixed
planes. In the normalization of \refs{\POLD, \TASI}, the RR charge of a
fixed plane is $-1$ while the charge of a four-brane is $1$ such that
cancellation holds globally. Local cancellation occurs in a
configuration with a four-brane at each orientifold
plane. The supersymmetric probes for this background are D0-branes whose
world-line effective action is expected to reproduce the string
background \DKPS . We will consider two distinct configurations:
\vskip 10pt

$\bullet$ $n$ D4-branes coalesce away from an orientifold fixed
plane. In $\doneneight$ language, the degrees of freedom on the
D0-brane world-line consist of an Abelian vector multiplet and an 
adjoint hypermultiplet arising from 0-0 strings and $n$
hypermultiplets in the fundamental of the $U(1)$ gauge group 
arising from 0-4 strings. The space-time positions of the four-branes
correspond to bare masses $\vec m_i$ of the charged multiplets in the gauge
theory on the probe. When the branes come together, one obtains
$SU(n)$ gauge symmetry enhancement in space-time corresponding to 
$SU(n)$ global symmetry enhancement in the probe theory.  

When the 0-brane is away from the 4-brane the massive 0-4 string
states can be integrated out. The surviving low energy degrees of
freedom in the world-line theory are the $\scrn=8$ vector multiplet
and neutral hypermultiplet. The later decouples so the low energy
effective action is the theory of an interacting $\scrn=8$ vector
multiplet. If the positions of the four-branes coincide the system is
rotationally invariant in the five transverse directions so the theory has
$Spin(5)\simeq Sp(2)$ symmetry under which the bosons transform in the
{\bf{5}} and fermions in the {\bf{4}}. 

The result of the previous 
section applies to this configuration and it remains to determine the
constants in \sol . In the present case,
$$
C^{\prime}={1\over g_s}
$$
is the asymptotic value of the dilaton far from the four-branes and at
the same time the classical coupling constant of the gauge theory on the
probe. The second constant $C$ is determined by the one-loop effects
of the $n$ charged hypermultiplets \DKPS\ to be 
$$
C=n.
$$
Therefore we conclude that the one-loop results of \DKPS\ are exact
already in this order and do not receive further corrections.
This statement is true as long as the theory is described in terms of the
multiplets introduced above but it may break down in a description in
terms of different variables. A similar phenomenon is encountered in three
dimensional gauge theories \refs{\SEIB, \SW, \IS} where the monopole
corrections become visible only after dualizing the photon. As we will see
latter, string duality suggests that this happens in the present case as well.

If the four-branes are localized at different points of coordinates
$\vec m_i$ in the transverse space, the $SU(n)$ global symmetry on the 
probe is broken since the hypermultiplets have different masses. In
this case one-loop metric is given by
\eqn\metthree{f(\vec x)={1\over g_s}+{1\over |\vec x-\vec m_1|^3}+\ldots+
{1\over |\vec x-\vec m_n|^3}.}
This configuration is no longer $Spin(5)$ symmetric in the transverse
directions, but the non-renormalization result still holds. The system
is still $\scrn=8$ supersymmetric so the exact $f$ must still satisfy
the five dimensional Laplace equation. The boundary conditions on the
exact metric close to $\vec m_i$ are given by
$$
f={1\over |\vec x-\vec m_i|^3},
$$
since near any of the $n$ D4 branes the remaining $n-1$ hypermultiplets
(corresponding to the the rest of the D4-branes) are very massive and
can be neglected. As $f$ is uniquely determined by the boundary
conditions the one-loop result \metthree\ is exact.

\vskip 2pt
$\bullet$ $n$ D4-branes coalesce at an orientifold fixed plane. The
degrees of freedom on the D0-brane consist now of a non-Abelian
$SU(2)$ vector multiplet plus an adjoint hypermultiplet arising from
0-0 strings and $n$ hypermultiplets in the fundamental of the gauge
group. The $n$ coalescing four-branes can be viewed as a collection of 
$2n$ branes pairwise identified by the $Z_2$ projection. Therefore
they are localized at points $\vec m_i,\,-\vec m_i$ in the transverse
space. The space-time gauge symmetry is enhanced to $SO(2n)$ when 
the branes coincide with the orientifold plane. As before this
corresponds to $SO(2n)$ global symmetry enhancement on the probe. 
The $SU(2)$ gauge group on the world-line is spontaneously broken to 
$U(1)$ by expectation values of the five scalars in the vector
multiplet which parameterize the Coulomb branch of the theory. 
Strictly speaking, this terminology is inappropriate as there is no real
moduli space in quantum mechanics. Nevertheless one can still refer
to a quantum mechanical moduli space in the Born-Oppenheimer
approximation \SEIB. 
In this sense the low energy effective action on the probe is a $U(1)$
gauge theory with $\scrn=8$ supersymmetry. When the four-branes
coalesce at the fixed plane there is a global $Spin(5)$ symmetry
rotating the five scalars in the  Abelian vector multiplet. Therefore
the general action is exactly of the form \fivelag . The only
difference with respect to the previous case is reflected in the value
of the constant $C$, 
$$
C= 2n-1
$$
where the negative term represents the one-loop contribution of
the non-Abelian vector multiplet to the effective action. As above 
there are no quantum corrections beyond one-loop. When the
four-branes are in general positions the one-loop metric is
$$
f(\vec x)={1\over g_s}+{1\over |\vec x-\vec m_1|^3}
+{1\over |\vec x+\vec m_1|^3}
+\ldots+{1\over |\vec x-\vec m_n|^3}
+{1\over |\vec x+\vec m_n|^3}-{1\over |\vec x|^3},
$$
and by the same argument used for \metthree\ does not get renormalized
beyond this order.
\newsec{Discussion}
 
We have shown above that the one dimensional action describing five bosons
and eight fermions in the {\bf 5} and {\bf 4} of $Spin(2)$ is, up to two
constants, uniquely determined by requiring $\scrn=8$ supersymmetry and
$Spin(5)$ invariance. Since the form of the action is fixed by a 
solution of a differential equation, we can not determine in our
formalism the constants that appear in the metric. Indeed, if the
probe is near an orientifold fixed plane with all four-branes far away
the metric becomes negative definite at a finite distance in moduli
space. The description of the physics in terms of the $\scrn=8$ vector
multiplet degrees of freedom breaks down and one has to look for
another set of variables. Similar phenomena occur in three
dimensional gauge theories where the equivalent description involves
dualizing the photon \refs{\SEIB, \SW, \IS}. In the new variables the three dimensional
non-renormalization theorem is violated by an infinite series of
monopole corrections \SW. 

This is also likely to be the case here since a dual set of variables
will not necessarily have a $Spin(5)$ symmetry. Further evidence for
this conclusion can be inferred from string duality arguments analogous
to those presented in \refs{\SEIB, \SW} for the three dimensional case.
The Type I' orientifold studied above is T-dual to Type I theory on a
five torus $T^5$ which is in turn S-dual to Heterotic string theory on the
same $T^5$. This is further dual to Type IIA theory on $K3\times S^1$ 
and after T-duality on the $S^1$ factor to Type IIB theory on
$K3\times\tilde S^1$. The D0-brane probe is mapped by the first
duality in the chain to a Type I D5-brane wrapped on $T^5$ while the 
D4-branes in the background are mapped to the $32$ D9-branes of Type I
theory. 
According to the analysis of \HAR\ the zero modes of the Type I
D5-brane wrapped on a four torus correspond to the world-sheet degrees
of freedom of the Type IIB string in static gauge. In our case the
five-brane is wrapped on an extra circle thus it maps to a fundamental
Type IIB string wrapped on the extra circle $\tilde S^1$ which is a
particle in the five non-compact dimensions. This is the image of the initial
D0-brane probe through the above chain of dualities. The resulting
sigma model is very different from the one we started with. 
It represents the motion of the particle on $K3\times \tilde S^1$, thus the
target space metric is the product of a hyper-K\"ahler metric on $K3$ and
a trivial metric on $S^1$. The fermions are target space vectors and
the symmetry is reduced to a product $U(1)\times G$ where $G$ is the
isometry group of the hyper-K\"ahler metric\footnote{$^*$}{The isometry group of a generic K3 surface is
trivial. However the moduli spaces of the probe theories are usually
non-compact pieces of the entire surface. In this case the hyper-K\"ahler
metric can have non-trivial isometry group.}. The 
orientifold background is mapped to a non-compact hyper-K\"ahler
manifold asymptotic to 
an $S^1$ bundle over the projective space $RP^2$. In particular
the metric is smooth and positive definite due to non-perturbative
corrections \SW. The singularities at infinite distance corresponding
four-brane backgrounds away are mapped to orbifold
singularities in the complex structure of the hyper-K\"ahler surface.

While in the three dimensional analysis of \refs{\SEIB, \SW}
the string duality picture is entirely reproduced by electric-magnetic
duality on a D2-brane probe, the present situation is less clear.
One could try to define an analogue of higher dimensional duality
transformations for the linear multiplet but we leave this for further study.
\bigskip
\centerline{\bf Acknowledgments}
We would like to thank T. Banks for suggesting the problem, as well as
for useful discussions. We are also grateful O. Aharony,
M. Berkooz, J. Gomis, N. Seiberg and E. Witten for discussions.

\appendix{A}{Spinor Conventions}

\subsec{$Sp(1)$ spinors}

The anticommuting coordinates $\theta_{\alpha}$ and
$\bart^{\alpha}\equiv(\theta_{\alpha})^*$ are spinors of $SU(2)\simeq Sp(1)$.
Raising and lowering indices is done with the $Sp(1)$ invariant metric as
\eqn\updown{\eqalign{\theta_{\alpha}&=\eps_{\alpha\beta}\theta^{\beta}
\hskip 20pt\theta^{\alpha}=\eps^{\alpha\beta}\theta_{\beta}\cr
{\bar{\theta}}_{\alpha}&=\eps_{\beta\alpha}{\bar{\theta}}^{\beta}\hskip 20pt{\bar{\theta}}^{\alpha}=\eps^{\beta\alpha}{\bar{\theta}}_{\beta}}.}
Complex conjugation of anticommuting numbers is defined by
\eqn\complex{(\eta_{\alpha}\xi_{\beta})^*={\bar{\xi}}^{\beta}{\bar{\eta}}^{\alpha,}}
which imply that $\partial_{\alpha}^*=-{\bar{\partial}}^{\alpha}$ and
$\partial^{{\alpha}*}=-{\bar{\partial}}_{\alpha}$. (This is necessary
for the solution $\sig$ of \consttwo\ to be consistent with the
reality condition \realone) . We also use
$$
\psi\psi\equiv\psi^\alpha\psi_\alpha,\hskip 20pt 
\overline{\psi\psi}\equiv\psb_\alpha\psb^\alpha,\hskip 20pt
\psi\psb\equiv\psi_\alpha\psb^\alpha=\psi^\alpha\psb_\alpha.
$$
The symmetric $\gamma$ matrices are
\eqn\gammathree{\sigma^1_{\alpha\beta}=i{\bf{1}}\hskip
20pt\sigma^2_{\alpha\beta}=\tau^3\hskip 20pt
\sigma^3_{\alpha\beta}=\tau^1,}
where $\tau$ are the Pauli matrices. They satisfy the algebra
\eqn\cliff{(\si\sj)_{\alpha\beta}=\delta^{ij}\eps_{\alpha\beta}
+i\eps^{ijk}\sigma^k_{\alpha\beta}}
 and the reality condition
\eqn\realtwo{(\si_{\alpha\beta})^*\equiv\sigma^{i\alpha\beta}=
\eps^{\alpha\gamma}\si_{\gamma\delta}\eps^{\delta\beta}.}

\subsec{$Sp(2)$ spinors}
We give the decomposition of $Sp(2)$ spinors and $\gamma$
matrices in terms of the corresponding $Sp(1)$ quantities which is
used to write the action and supersymmetry variations in a $Spin(5)$
form. Unless otherwise noted, all conventions are similar to those
used above. Written in terms of $Sp(1)$ spinors, the $Sp(2)$ ones are 
\eqn\spinors{\eta_\alpha=\left(\matrix{\lambda_\alpha\cr\psb_\alpha}\right)
\hskip 20pt\bar\eta^\alpha=\left(\matrix{\lb^\alpha\cr\psi^\alpha}\right).}
Indices are raised lowered and contracted using the metric
\eqn\invmet{J_{\alpha\beta}=\pmatrix{0&\eps\cr\eps&0},}
with $\eps$ being the $Sp(1)$ metric. The antisymmetric $\gamma$
matrices are taken to be
\eqn\matrices{\eqalign{\gamma^i_{\alpha\beta}&=\pmatrix{0&\si\cr-\si&0}\hskip
10pt i=1,2,3\cr
\gamma^4_{\alpha\beta}&=\pmatrix{i\eps&0\cr0&-i\eps}\cr
\gamma^5_{\alpha\beta}&=\pmatrix{-\eps&0\cr0&-\eps},}}
with the reality condition being
\eqn\realthree{(\gamma^1_{\alpha\beta})^*\equiv\gamma^{i\,\alpha\beta}=-J^{\alpha\gamma}\gamma^i_{\gamma\delta}J^{\delta\beta}.}

\appendix{B}{Lagrangian and SUSY variations}

$\bullet$ $\scrn=4$ lagrangian
\eqn\superspace{\eqalign{
&{1\over{4}}\kii\left(\dot{x}^j\dot{x^j}+i\left(\lb\dot{\lambda}+\lambda\dot{\lb}\right)+D^2\right)
-\kff\left(\dot{\Phi}\dot{\fb}+i\left(\psb\dot{\psi}+\psi\dot{\psb}\right)+FF^*\right)\cr
&+{1\over{2}}\dot{x}^j\left(\kiif\psi\sj\lambda+\kiib\psb\sj\lb\right)
+\dot{x}^i\eps^{ijk}\left(-{1\over{4}}\kllk\lambda\sj\lb+\kffk\psi\sj\psb\right)\cr
&+i\dot{\fb}\left({1\over{4}}\kiib\lambda\lb-\kffb\psi\psb-i\kffi\psi\si\lambda\right)\cr
&-i\dot{\Phi}\left({1\over{4}}\kiif\lambda\lb-\kfff\psi\psb+i\kffi\psb\si\lb\right)\cr
&+D\left({1\over{4}}\klli\lambda\si\lb+\kffi\psi\si\psb
+{i\over{2}}\left(\kiif\psi\lambda+\kiib\overline{\psi\lambda}\right)\right)\cr
&+F\left(-{1\over{4}}\kiif\lambda\lambda-\kffb\overline{\psi\psi}+i\kffi\lambda\si\psb\right)\cr
&+F^*\left({1\over{4}}\kiib\overline{\lambda\lambda}+\kfff\psi\psi-i\kffi\psi\si\lb\right)}}
$$\eqalign{
&+{1\over{4}}\left(\kiibb\overline{\lambda\lambda}\,\overline{\psi\psi}+\kiiff\lambda\lambda\,\psi\psi\right)+\kffff\overline{\psi\psi}\,\psi\psi\cr
&+{1\over{16}}\kllii\overline{\lambda\lambda}\,\lambda\lambda-\kffij\psi\si\lb\,\lambda\sj\psb\cr
&-{i\over{4}}\left(\kiijb\lambda\sj\psb\,\overline{\lambda\lambda}+\kiijf\psi\sj\lb\,\lambda\lambda\right)\cr
&-i\left(\kffif\lambda\si\psb\,\psi\psi+\kffib\psi\si\lb\,\overline{\psi\psi}\right)}
$$
$\bullet$ $\scrn=8$ lagrangian\footnote{$^*$}{There is a sign
ambiguity in expressions of the form $\psi\si\lambda$ or
$\psb\si\lb$, corresponding to lower or upper
indices on the spinors. We universally take spinors with lower
indices. No such ambiguity arises if one of the spinors is barred.} --
Follows form the $\scrn=4$ one by making use of \mettwo .
\eqn\componentlag{\eqalign{{\cal{L}}=&-f\left(\dot{x}^i\dot{x}^i+\dot{\Phi}\dot{\fb}+i({\lb}{\dot{\lambda}}+\lambda{\dot{\lb}}+
{\psb}{\dot{\psi}}+\psi{\dot{\psb}})\right)\cr
+&\dot{x}^if_{,k}\eps^{ijk}(\psi\sj{\psb}+\lambda\si{\lb})
-2\dot{x}^i\left(f_{,\Phi}\psi\si\lambda+f_{,\fb}\psb\si\lb\right)\cr
+&\dot{\Phi}\left(f_{,i}\psb\si\lb+if_{,\Phi}\psi\psb
+if_{,\fb}\lambda\lb\right)
+\dot{\fb}\left(f_{,i}\psi\si\lambda-if_{,\Phi}\psi\psb
-if_{,\fb}\lambda\lb\right)\cr
+&if_{,i\Phi}\left(\lambda\si\psb\,\,\overline{\lambda\lambda}
-\psi\si\lb\,\,\overline{\psi\psi}\right)
+if_{,i\fb}\left(\psi\si\lb\,\,\lambda\lambda
-\lambda\si\psb\,\,\psi\psi\right)\cr
-&f_{,\Phi\Phi}\lambda\lambda\,\,\psi\psi
-f_{,\fb\fb}\overline{\lambda\lambda}\,\,\overline{\psi\psi}
-f_{,ij}\lambda\si\psb\,\,\psi\si\lb
+f_{,\Phi\fb}\psi\psi\,\,\overline{\psi\psi}
-{1\over{4}}f_{,ii}\lambda\lambda\,\,\overline{\lambda\lambda}\cr
-&fD^2+D\left(f_{,i}\left(\psi\si\psb-\lambda\si\lb\right)
-2f_{,\Phi}\psi\lambda-2f_{,\fb}\overline{\psi\lambda}\right)\cr
-&fFF^*+F\left(f_{,\Phi}\lambda\lambda-f_{,\fb}\overline{\psi\psi}+if_{,i}\lambda\si\psb\right)-F^*\left(f_{,\fb}\overline{\lambda\lambda}-f_{,\Phi}\psi\psi+if_{,i}\psi\si\lb\right)
}}
$\bullet$ Auxiliary fields -- These are given $Spin(5)$ language
using the results of appendix A.
\eqn\aux{\eqalign{D&={1\over{2}}f^{-1}f_{,i}\,\eta^\alpha\gamma^i_{\alpha\beta}\bar\eta^\beta\cr
F&={i\over{2}}f^{-1}f_{,i}\,\bar\eta^\alpha\gamma^i_{\alpha\beta}\bar\eta^\beta\cr
F^*&={i\over{2}}f^{-1}f_{,i}\,\eta^\alpha\gamma^i_{\alpha\beta}\eta^\beta
}}
$\bullet$ Manifest SUSY
\eqn\manvar{\eqalign{\delta
x^i&=-i\left(\eps\si\lb+\epb\si\lambda\right)\cr
\delta\lambda_\alpha&=\dot{x}^i(\eps\si)_\alpha+iD\eps_\alpha
\hskip 20pt
\delta\lb_\alpha=\dot{x}^i(\epb\si)_\alpha-iD\epb_\alpha\cr
\delta D&=\eps\dot{\lb}-\epb\dot{\lambda}\cr
\delta\Phi&=2\eps\psi\hskip
79pt\delta\bar\Phi=-2\overline{\eps\psi}\cr
\delta\psi_\alpha&=-i\dot{\Phi}\epb_\alpha+F\eps_\alpha\hskip
32pt\delta\bar\psi_\alpha=-i\dot{\fb}\eps_{\alpha}+F^*\epb_\alpha\cr
\delta F&=-2i\epb\dot\psi\hskip 62pt\delta F^*=-2i\eps\dot{\bar\psi}}
}
$\bullet$ Non-manifest SUSY
\eqn\nonmanvar{\eqalign{
\delta x^i&=-i\left(\eps\si\psi-\epb\si\bar\psi\right)\cr
\delta\lambda_\alpha&=i\dot{\Phi}\eps_\alpha+F^*\epb_\alpha\hskip 39pt
\delta\lb_\alpha=-i\dot{\fb}\epb_\alpha-F\eps_\alpha\cr
\delta D&=\eps\dot{\psi}-\overline{\eps\dot{\psi}}\cr
\delta\Phi&=-2\epb\lambda\hskip
76pt\delta\bar\Phi=2\eps\lb\cr
\delta\psi_\alpha&=\dot{x}^i(\epb\si)_\alpha+iD\epb_\alpha\hskip
24pt\delta\bar\psi_\alpha=-\dot{x}^i(\eps\si)_{\alpha}+iD\eps_\alpha\cr
\delta F&=-2i\overline{\eps\dot\lambda}\hskip 67pt\delta F^*=-2i\eps\dot\lambda.}}
$\bullet$ $Spin(5)$ SUSY -- These follow from \manvar\ and \nonmanvar\
using \aux\ and the results of appendix A.
\eqn\susyfive{\eqalign{\delta
x^i&=i\left(\eps^\alpha\gamma^i_{\alpha\beta}\bar\eta^{\beta}-\epb^{\alpha}\gamma^i_{\alpha\beta}\eta^\beta\right)\cr
\delta\eta_\alpha&=\dot{x}^i\gamma^i_{\alpha\beta}\eps^{\beta}+{i\over{2}}f^{-1}f_{,i}\left(\eta^\gamma\gamma^i_{\gamma\delta}\bar\eta^\delta\,\,\eps_\alpha
+\eta^\gamma\gamma^i_{\gamma\delta}\eta^\delta\,\epb_\alpha\right)\cr
\delta\bar\eta_\alpha&=-\dot{x}^i\gamma^i_{\alpha\beta}\epb^{\beta}-{i\over{2}}f^{-1}f_{,i}\left(\eta^\gamma\gamma^i_{\gamma\delta}\bar\eta^\delta\,\,\epb_\alpha
+\bar\eta^\gamma\gamma^i_{\gamma\delta}\bar\eta^\delta\,\eps_\alpha\right)
.}}

\vfill
\listrefs

\end